\documentclass[twocolumn]{aastex701}

\newcommand{\kmpspmpc}{\,kms$^{-1}$Mpc$^{-1}$}
\newcommand{\TPnineteen}{\ensuremath{T_{\mathrm{inf,\:P19}\:}}}

\newcommand{\TPz}{\ensuremath{T_{\mathrm{inf,\:z=0}}}}
\newcommand{\V}{$V_{\text{max}}$}
\newcommand{\alphacold}{$\alpha_{\mathrm{cold}}\,$}
\newcommand{\alphahot}{$\alpha_{\mathrm{hot}}\,$}
\newcommand{\SHARKspec}{{\sc Shark}$_{\mathrm{spec}}\,$}
\defcitealias{Oxland2024}{O24}
\shorttitle{Satellite Quenching in Shark v2.0}
\shortauthors{Oxland et al.}

\begin{document}

\title{The role of gas stripping in the quenching of satellite galaxies using SHARK v2.0}

\author[orcid=0009-0003-2618-5955,gname=Megan, sname=Oxland]{Megan K. Oxland}
\affiliation{Department of Physics and Astronomy, McMaster University, 1280 Main Street West, Hamilton, ON, L8S 4L8, Canada}
\email[show]{oxlandm@mcmaster.ca}  

\author[orcid=0000-0001-5742-7927,gname=Matias, sname=Bravo]{Mat\'ias Bravo} 
\affiliation{Department of Physics and Astronomy, McMaster University, 1280 Main Street West, Hamilton, ON, L8S 4L8, Canada}
\email{bravosam@mcmaster.ca}

\author[orcid=0000-0003-4722-5744, gname=Laura, sname=Parker]{Laura C. Parker}
\affiliation{Department of Physics and Astronomy, McMaster University, 1280 Main Street West, Hamilton, ON, L8S 4L8, Canada}
\email{lparker@mcmaster.ca}

\author[orcid=0000-0003-3021-8564,gname=Claudia, sname=Del P. Lagos]{Claudia Del P. Lagos} 
\affiliation{International Centre for Radio Astronomy Research, The University of Western Australia, 35 Stirling Highway, Crawley, Western Australia 6009, Australia}
%\affiliation{ARC Centre for Excellence for All-Sky Astrophysics in 3 Dimensions (ASTRO 3D)}
\affiliation{Cosmic DAWN Center, Niels Bohr Institute, University of Copenhagen, Jagtvej 128, Copenhagen N, DK-2200, Denmark}
\email{claudia.lagos@icrar.org}

\begin{abstract}
Observational studies have made substantial progress in characterizing quenching as a function of stellar mass and environment, but they are often limited in their ability to constrain quenching timescales and to determine the dominant environmental process responsible for the shutting down of star formation. To address this, we combine recent Sloan Digital Sky Survey (SDSS) observations with the {\sc Shark} v2.0 semi-analytic model to study the quenching of satellite galaxies in groups and clusters. We generate mock SDSS-like observations to calibrate the hot halo and cold interstellar medium (ISM) gas stripping prescriptions against observed satellite quenched fractions, finding that the previously adopted stripping prescriptions in {\sc Shark} v2.0 are too aggressive and overestimate the quenched fraction of satellite galaxies. Reducing the efficiency of both hot and cold gas stripping yields excellent agreement with observations for low- and intermediate-mass satellite galaxies. We use the calibrated model to investigate quenching timescales and find that satellites quench more quickly in clusters compared to groups, with timescales that generally decrease with increasing stellar mass. The long ($>2$ Gyr) timescales we measure favour hot halo gas removal as the dominant driver of satellite quenching.
\end{abstract}


\section{Introduction}
\label{sec:intro}
Galaxy evolution is driven by a combination of internal and environmental processes that regulate and eventually shut down star formation \citep[e.g.][]{Peng2010, Cortese2021}. This transformation is reflected in the bimodal distribution of galaxy colours and star formation rates in the local universe \citep[e.g.,][]{Strateva2001, Baldry2004, Wetzel2012}, where most galaxies populate either the ``blue cloud" or ``red sequence", with relatively few found in the transitional``green valley" region \citep{Salim2014, Schawinski2014}. While a small subset of passive galaxies may undergo rejuvenation and resume star formation \citep[e.g.,][]{Thomas2010, Cleland2021}, the vast majority experience quenching marked by a long-term decline in their star formation rates (SFRs). In dense environments, this transition is often accompanied by systematic morphological changes from late-type, disc-dominated systems to early-type, bulge-dominated ones as evidenced by the morphology-density relation \citep[e.g.,][]{Dressler1980} highlighting the strong role of environment in shaping galaxy properties. The quenched fraction of galaxies scales with both stellar mass and host halo mass \citep[e.g.,][]{Peng2010, Wetzel2012}, further supporting a picture in which both internal and environmental processes drive galaxy evolution. 

All galaxies are subject to internal quenching mechanisms. Feedback from supernovae and outflows driven by active galactic nuclei (AGN) can inject large amounts of energy into the surrounding medium, reducing the supply of fuel available for star formation \citep[e.g.,][]{Croton2006, Bower2006, Fabian2012}. Galactic bars can transport gas into the central regions, triggering a central starburst and the subsequent growth of a bulge component \citep{Kormendy2004}. A large bulge can stabilize the gas disc against fragmentation, suppressing star formation through morphological quenching \citep{Martig2009}. Galaxies residing in dense environments are also subject to external quenching mechanisms \citep[see review by][]{Cortese2021}. Ram pressure stripping (RPS) exerted by the dense intracluster medium (ICM) can remove a galaxy's hot halo gas, cutting off the supply of material that would have eventually cooled and replenished the interstellar medium \citep[ISM, a process often referred to as starvation; e.g.,][]{Larson1980, Peng2015}. In more extreme cases, RPS can also strip the cold ISM itself, directly depleting the reservoir of star-forming gas \citep[e.g.,][]{Gunn1972}. Repeated close encounters with neighbouring galaxies \citep{Moore1996} and tidal interactions \citep{Gnedin2003} can further remove material, impacting subsequent star formation. Mergers can trigger starbursts and AGN, leading to the eventual quenching of the galaxy \citep[e.g.,][]{Hopkins2006, Ellison2013}.

While the physical mechanisms described above are well understood, there remains no clear consensus as to the dominant environmental mechanisms responsible for quenching satellite galaxies. Previous observational studies have highlighted the roles of RPS, starvation, tidal interactions, and harassment, yet their relative importance has been shown to depend on factors such as stellar mass, halo mass, orbital history, redshift, and pre-processing \citep[e.g.,][]{vandenBosch2008, Wetzel2013, Hatfield2017, Webb2020, Singh2025}. Crucially, these mechanisms operate on distinct timescales, meaning that the rate at which galaxies quench provides a diagnostic of their relative importance. In addition, timescales can help us understand whether satellite quenching proceeds rapidly or after some delay time, which is still an open question \citep[e.g.,][]{Balogh2000, Wetzel2013, Belli2019, Rhee2020, Oman2021, Reeves2023}. 

Previous work by \citet[hereafter \citetalias{Oxland2024}]{Oxland2024} looked at a sample of group and cluster galaxies from the Sloan Digital Sky Survey (SDSS; \citealt{Abazajian2009}) to study how star formation rates change over time for infalling satellite galaxies. They found the quenched fraction increases with time since infall, with the strongest trends found for low mass galaxies falling into high mass clusters. Their analysis favoured RPS and starvation as key drivers, but was unable to disentangle the dominant mechanism observationally. A natural follow up to this work is to use simulations, calibrated with these observations, to study the quenching timescales and physical mechanisms responsible for the transformation of satellite galaxies.

Semi-analytic models of galaxy formation (SAMs) are a class of simulation that solve a set of coupled differential equations describing the key physical processes involved in galaxy formation and evolution, applied to merger trees extracted from dark matter-only $N$-body simulations \citep{Baugh2006, Benson2010}. Compared to hydrodynamical simulations, SAMs are computationally inexpensive, allowing for the testing of different physical prescriptions and the efficient exploration of large parameter spaces. Over the past two decades, SAMs have been successful in reproducing observations including the luminosity function, galaxy colour distributions, stellar mass functions, passive fractions, and the cosmic star formation history \citep[e.g.][]{Croton2006, Bower2006, Lagos2008, Cora2018, Guo2011, Henriques2015}. For satellites, SAMs have demonstrated their ability to match observed quenched fractions, but only when gradual stripping is implemented \citep[e.g][]{Font2008, Kang2008, Cora2018, Xie2020}. Gradual stripping has recently been implemented in {\sc Shark}, the SAM used in this work, yet its effect has not been directly compared to observational studies.

We use the SAM {\sc Shark} \citep{Lagos2018, Lagos2024} to study how satellite galaxies quench in dense environments, first calibrating {\sc Shark}'s gas stripping prescriptions using the observational results of \citetalias{Oxland2024}. We then use the calibrated model to explore the roles of hot and cold gas stripping in satellite quenching. This paper is organized as follows. In Section \ref{sec:obs}, we briefly describe the observations of \citetalias{Oxland2024} which we use to calibrate the SAM {\sc Shark}, introduced in Section \ref{sec:sim}. Section \ref{sec:mockobs} details the methods used to generate mock SDSS-like observations of {\sc Shark} galaxies. Section \ref{sec:calibration} presents our calibration of the {\sc Shark} environmental stripping prescriptions, and Section \ref{sec:satellite_quenching} uses this calibration to study the quenching of satellite galaxies in {\sc Shark}. We discuss our results, biases and caveats in Section \ref{sec:discussion}, and summarize our main conclusions in Section \ref{sec:conclusions}.

%%%%%%%%%%%%%%%%%%%%%%%%%%%%%%%%%%%%%%%%%%%%%%%%%%

\section{Observational SDSS Sample}
\label{sec:obs}
Here we summarize the observational data used in this work, but refer the reader to \citetalias{Oxland2024} for a more thorough description of their sample. In brief, \citetalias{Oxland2024} selected low redshift galaxies ($z\leq0.1$) from the SDSS Data Release 7 \citep{Abazajian2009}, and adopted the Yang Group Catalogue \citep{Yang2007} to determine group membership. Star formation rates and stellar mass measurements were obtained from the GALEX-SDSS-WISE Legacy Catalogue \citep{Salim2016,Salim2018}. A volume correction of $1/V_{\mathrm{max}}$ was applied to the quenched fractions to correct for the Malmquist bias, with $V_{\mathrm{max}}$ values taken from the catalogue of \cite{Simard2011}.  

The sample was restricted to galaxies with stellar masses $\log_{10}(M_{\star}/\mathrm{M}_{\odot}) \geq 9.5$, residing in groups ($13 \leq \log_{10}(M_{\text{halo}}/\mathrm{M}_{\odot}) < 14$) or clusters ($\log_{10}(M_{\text{halo}}/\mathrm{M}_{\odot}) \geq 14$) containing at least three members. The most massive galaxy in each group/cluster (as identified by \citealt{Yang2007}) was excluded to remove central galaxies. Their final sample contains 20369 satellites in total, with 12062 in groups and 8307 in clusters, along with a control sample of 73474 isolated field galaxies. This SDSS sample assumes flat, $\Lambda$CDM cosmology with $\Omega_M = 0.3$, $\Omega_\Lambda = 0.7$, and $H_0 = 70$\kmpspmpc. This is the observational sample we attempt to replicate with {\sc Shark} in Section \ref{sec:mockobs}.

%%%%%%%%%%%%%%%%%%%%%%%%%%%%%%%%%%%%%%%%%%%%%%%%%%

\section{The {\sc Shark} Semi-Analytic Model}
\label{sec:sim}
{\sc Shark} is an open-source\footnote{Available on GitHub: \href{https://github.com/ICRAR/shark}{\url{https://github.com/ICRAR/shark}}. Static version used in this work: \citet{sharkv2_o25}}, highly modular SAM of galaxy formation first introduced as {\sc Shark} v1.0 in \cite{Lagos2018}. As with most SAMs, {\sc Shark} solves a set of coupled differential equations that model the most critical physical processes in galaxy formation and evolution including (i) the formation and collapse of dark matter (DM) haloes; (ii) gas accretion onto DM haloes; (iii) the subsequent cooling of said gas, leading to the formation of galaxy discs via the conservation of specific angular momentum; (iv) star formation in galaxy discs; (v) stellar feedback and subsequent chemical enrichment of the ISM from evolving stars; (vi) disc instabilities transferring gas and/or stars into the central bulge component; (vii) the growth of supermassive black holes (SMBHs) from gas accretion and SMBH mergers; (viii) AGN feedback; (ix) photo-ionization of the intergalactic medium; (x) the growth of galaxy spheroids and the triggering of starbursts via disc instabilities and galaxy mergers; (xi) environmental processes affecting satellite galaxies \citep{Lagos2024}. The best fit parameters governing these models are typically calibrated against observations.

Galaxies in {\sc Shark} are classified as either centrals ({\tt type=0}), satellites ({\tt type=1}), or orphans ({\tt type=2}) depending on the properties of the halo in which they inhabit. All galaxies start off as centrals in isolated DM haloes. During mergers of haloes, galaxies may either remain a central if belonging to the main progenitor halo, or conversely become a satellite of a subhalo embedded within the central host halo. Once a subhalo can no longer be tracked due to poor resolution, the galaxy becomes an orphan and its position is randomly drawn from a 3D Navarro–Frenk–White (NFW) profile with the same properties of its host halo. Orphan galaxy velocities are also drawn from an NFW profile, assuming isotropic velocities \citep[see Section 4.4.7 of][]{Lagos2018}. The orphans eventually merge with the central on a dynamical friction timescale \citep{Lagos2018, Lagos2024}. Orphans and satellites are subject to environmental effects, while centrals are not. 

{\sc Shark} has been tested extensively and reproduces a variety of observations including galaxy luminosity functions from the far-ultraviolet to far-infrared \citep{Lagos2019}, the optical colour bimodality \citep{Bravo2020}, the atomic hydrogen-halo mass relation \citep{Chauhan2021}, the SFMS and its scatter \citep{Davies2019} and the number counts of sub-millimetre galaxies \citep{Lagos2020}. Despite these successes, a few tensions remained such as the limited number of massive quiescent galaxies at $z\gtrsim2$ \citep{Long2023, Gould2023, Lagos2025} and the quenching timescales of {\sc Shark} galaxies being too short compared to observations \citep{Bravo2023}. These tensions inspired the latest update, {\sc Shark} v2.0 \citep{Lagos2024}, which aimed to improve the modelling of galaxy evolution over cosmic time. The most significant improvements over the first version of {\sc Shark} include an improved modelling of the momentum exchange between the ISM and stars, improved AGN feedback that now includes both wind and jet feedback modes, a model tracking black hole spins, and improved modelling of the environmental effects impacting satellite galaxies. While satellite galaxies in {\sc Shark} v1.0 experienced instantaneous stripping of their hot halo (with no cold gas stripping), {\sc Shark} v2.0 now includes separate prescriptions for the gradual stripping of both the hot halo and cold ISM gas (see Section \ref{gas_strip}), as well as tidal stripping of baryonic material. These improvements, in particular the updated environmental modelling, warrant a deeper examination of the quenching of satellite galaxies in {\sc Shark}.

{\sc Shark} is run over the SURFS suite of N-body, DM only simulations of \citet{Elahi2018}. These simulations adopt the $\Lambda$CDM \citet{Planck2016} cosmology, with $\Omega_M = 0.3121$, $\Omega_\Lambda = 0.6879$, and $H_0 = 67.51$\kmpspmpc. The simulation used here is the L210N1536 box, which has a side length of $210 h^{-1}$Mpc, with $1536^{3}$ DM particles each with masses of $2.21\times10^8 h^{-1}\mathrm{M}_{\odot}$. This box was recently rerun using the {\sc SWIFT} $N$-body and Smooth Particle Hydrodynamics solver \citep{Schaller2024}. From this, halo catalogues and merger trees were built using HBT-HERONS \citep{ForouharMoreno2025, ChandroGomez2025}, which can be done on-the-fly in SWIFT. HBT-HERONS improves upon previous halo finders which have been shown to suffer from numerical issues in dense environments including central/satellite mass swapping events and the sudden appearance of massive subhaloes \citep{ChandroGomez2025}. 

\subsection{Environmental Stripping of Satellite Galaxies}
\label{gas_strip}
As noted above, {\sc Shark} v2.0 has an improved treatment of environmental effects with separate prescriptions for the gradual stripping the hot halo and cold ISM gas of satellite galaxies. We summarize both below, and refer the reader to Section 3.4 in \citet{Lagos2024} for full details. 

The hot halo gas stripping follows \citet{Font2008} where the gas beyond a radius of $r_{\mathrm{sat}}$ (measured from the centre of the satellite galaxy) is removed if the ram pressure at the satellite's position in the host halo exceeds the gravitational binding energy of its halo gas. Specifically, the hot halo gas stripping prescription is
\begin{equation}
    \rho^{\mathrm{cen}}_{\mathrm{halo, gas}}(R)\, v^2_{\mathrm{sat}} > \alpha_{\mathrm{hot}} \, \frac{G \, M_{\mathrm{sat}}(r_{\mathrm{sat}}) \, M^{\mathrm{sat}}_{\mathrm{halo, gas}}}{8 \, r^{\mathrm{sat}}_{\mathrm{vir}} \, r^3_{\mathrm{sat}}},
    \label{hot_eqn}
\end{equation}
where $\rho^{\mathrm{cen}}_{\mathrm{halo, gas}}(R)$ is the density of the central galaxy's host halo at the satellites position relative to the central, and $v_{\mathrm{sat}}$ is the relative velocity between the central and satellite. $M_{\mathrm{sat}}(r_{\mathrm{sat}})$ is the mass enclosed within $r_{\mathrm{sat}}$, and $M^{\mathrm{sat}}_{\mathrm{halo, gas}}$ and $r^{\mathrm{sat}}_{\mathrm{vir}} $ are the total gas mass and the virial radius of the satellite immediately before it became a satellite, respectively. Gas is removed outside-in, leaving the density profile inside the stripping radius unchanged. \alphahot is a free parameter, set to one by default, that governs the strength of the hot gas stripping. In {\sc Shark}, this is controlled by {\tt alpha\_rps\_halo} in the parameter file.

Similarly, any cold ISM gas outside of a radius $r$ is removed if 
\begin{equation}
    \rho^{\mathrm{cen}}_{\mathrm{halo, gas}}(R) \, v^2_{\mathrm{sat}} > \alpha_{\mathrm{cold}} \ 2 \pi G \, \Sigma_{\mathrm{gas}}(r) \left[ \Sigma_{\mathrm{gas}}(r) + \Sigma_{\star}(r) \right]
    \label{cold_eqn}
\end{equation}
is satisfied, which comes from \citet{Tecce2010} based on the simple criterion proposed by \citet{Gunn1972}. $\Sigma_{\mathrm{gas}}(r)$ and $\Sigma_{\mathrm{\star}}(r)$ are the ISM and stellar surface densities at $r$, respectively. \alphacold is a free parameter we have added to {\sc Shark}, set to $1$ by default, to control the strength of cold gas stripping. It is set by {\tt alpha\_cold} in the parameter file. {\sc Shark} imposes a minimum hot gas stripping radius of $r^{\mathrm{sat}}_{\mathrm{vir}}/100$ and a minimum cold gas stripping radius of $r^{\mathrm{sat}}_{\mathrm{vir}}/500$ to prevent numerical issues at small radii. Since  \alphahot and \alphacold are directly related to the amount of hot halo and cold ISM gas that is removed, they serve as parameters that control how fast galaxies quench by starvation and RPS, respectively. 

\begin{figure}
    \includegraphics[width=\linewidth]{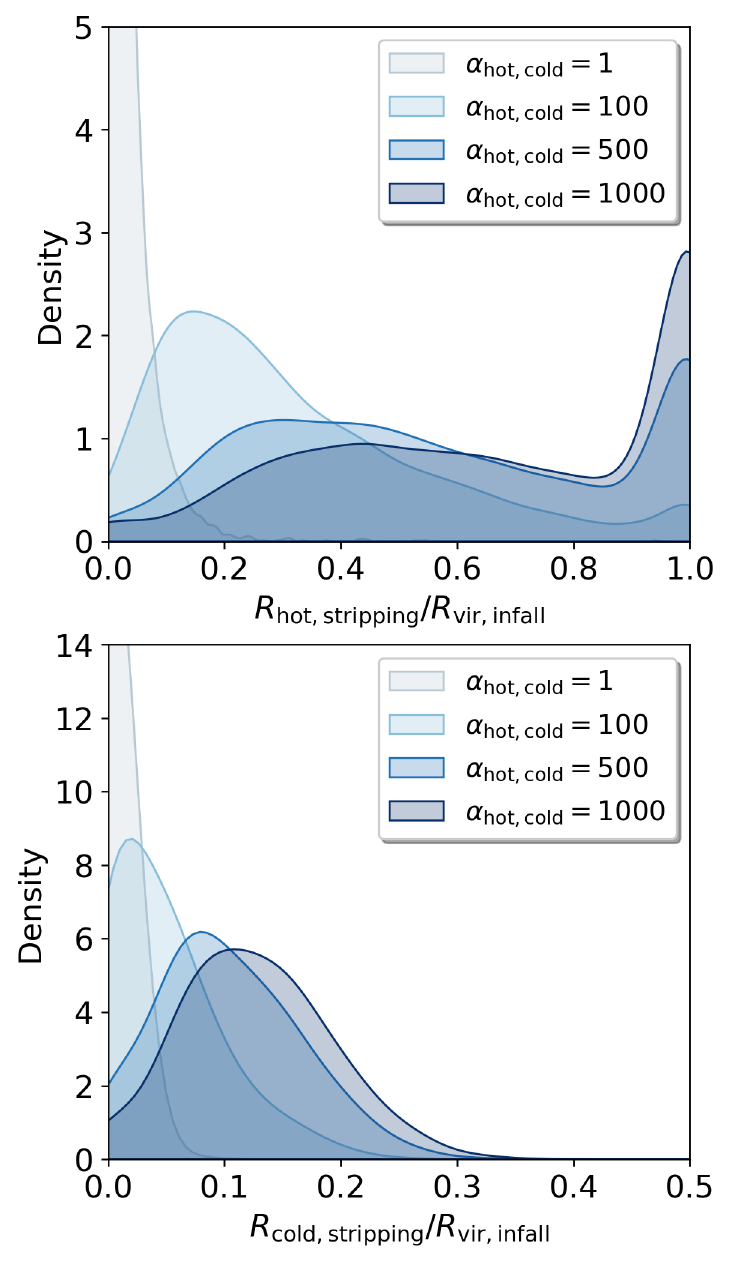}
    \caption{Distribution of hot halo gas stripping radii (top panel) and cold ISM gas stripping radii (bottom panel) normalized by the virial radius of the galaxy at infall for satellites with $9 < \log_{10}(M_{\star}/\mathrm{M}_{\odot}) < 9.5$ falling into clusters. Different colours correspond to {\sc Shark} runs with different  \alphahot and \alphacold parameters, as shown in the legend. Note the top and bottom panels have different limits on the x-axis, and the y-axis is truncated to more clearly show the differences between different \alphahot and \alphacold runs.}
    \label{fig:strpping_radii}
\end{figure}

It is important to clarify the role of \alphahot and \alphacold in gas stripping. These free parameters appear on the right-hand side of Equations \ref{hot_eqn} and \ref{cold_eqn}, where they effectively change the binding energy of the gas. As such, they are inversely related to the efficiency of hot and cold gas stripping: increasing \alphahot or \alphacold raises the gas binding energy, which in turn reduces the amount of gas that can be stripped by increasing the stripping radii. This effect can be seen in Figure \ref{fig:strpping_radii} where we show the distribution of hot halo gas and cold ISM gas stripping radii (normalized by the virial radius the galaxy had at infall) for the subset of galaxies with $9 < \log_{10}(M_{\star}/\mathrm{M}_{\odot}) < 9.5$ falling into clusters with different values of \alphahot and \alphacold. Increasing \alphahot and \alphacold shift the distributions toward larger stripping radii, with the effect more pronounced for hot halo gas (top panel) than from cold ISM gas (bottom panel). At large \alphahot, the distribution of galaxies peaks near $\mathrm{R_{vir, infall}}$ indicating no removal of the hot halo gas. Cold stripping radii are much smaller, only reaching $\sim 0.3 \:\mathrm{R_{vir, infall}}$, as the cold gas is contained in the disc and its radial extent is much smaller than that of the hot halo.  

The removal of hot halo and cold ISM gas are implemented as separate processes. The hot gas stripping radius determines the amount of hot halo gas to be removed, while the the cold gas stripping radius determines how much cold ISM should be removed. These radii are recalculated at every snapshot and do not directly depend on one another. However, because gas can flow between the bulge, disk, and halo components, the two processes are indirectly coupled.

From Equations \ref{hot_eqn} and \ref{cold_eqn} we find \alphahot and \alphacold scale with the stripping radius as $r\propto\alpha_{\mathrm{hot}}^{1/3}$ and $r\propto\alpha_{\mathrm{cold}}^{1/4}$. These scalings should be kept in mind when interpreting the best-fit values of \alphahot and \alphacold presented in Section \ref{sec:calibration}, as the effect on the stripping radii is small compared to the numerical values of \alphahot and \alphacold chosen.

In addition to hot and cold gas stripping, {\sc Shark} v2.0 implemented a model for the tidal stripping of stellar material (see Section 3.5 of \citealt{Lagos2024}). If there is a cold gas disc, any gas beyond where stars have been stripped will also been removed. In principle, the tidal stripping radii are larger than the RPS radii for satellite galaxies, so there is often no cold ISM gas left to strip. Orphan galaxies are affected by tidal stripping by construction, however with the new HBT-HERONS merger trees there are very few orphan galaxies. This model is on by default, but can be turned off using {\tt tidal\_stripping} in the {\sc Shark} parameter file. We keep tidal stripping on in our analysis, but tested with it switched off and can confirm that our results remain unchanged.

\subsection{Improvements in {\sc Shark} v2.0}
\label{improvements}
Since the release of {\sc Shark} v2.0 \citep{Lagos2024}, a number of minor changes have been made to the code to better match observations. Two of these changes have already been introduced in this paper: the HBT-HERON subhalo finder/merger tree builder used (see Section \ref{sec:sim}) and the addition of the free parameter \alphacold that controls the strength of cold gas stripping (see Section \ref{gas_strip}). Additional changes include (i) modifying the time dependence of stellar feedback in the \citet{Lagos2013} model, replacing the (1+z) dependence with a dependence on the age of the universe; and (ii) introducing a redshift threshold for AGN feedback below which there is a memory of the AGN heating radius. This is a free parameter, set to $1.669$ for this work. 

The change in stellar feedback was done to allow galaxies in the early universe to have a more slowly evolving stellar feedback efficiency. This change only produces a visible impact on the predictions of the model at $z\gtrsim 3$. For the change in AGN feedback, we note that the lack of memory of the AGN heating radius makes AGN feedback slightly less efficient above the redshift at which memory is turned on. This allows a faster build-up of mass for massive galaxies early on, but again this change makes a difference only in the early universe $z\gtrsim 3$. With these changes, the parameters of the model were re-tuned to reproduce the stellar mass function of galaxies at $z=0$~and~$=2$, following the process described in Section~3.6 of \citet{Lagos2024}.

%%%%%%%%%%%%%%%%%%%%%%%%%%%%%%%%%%%%%%%%%%%%%%%%%%

\section{Generating Mock Observations}
\label{sec:mockobs}
To accurately calibrate \alphahot and \alphacold in Equations \ref{hot_eqn} and \ref{cold_eqn} using the results of \citetalias{Oxland2024}, we generate mock observations of our {\sc Shark} galaxies that match the selection function of the SDSS as closely as possible. The first step in this process is to generate a light-cone (LC) from the {\sc Shark} outputs, with comparable sky and redshift coverage to the SDSS. To produce the LC, we use the publicly available code STINGRAY\footnote{Available on GitHub:\href{https://github.com/obreschkow/stingray}{\url{https://github.com/obreschkow/stingray}}. Static version: \citet{stingray_043}}, which is an updated version of the LC generating code by \cite{Obreschkow2009} described in \citet{Chauhan2019}. Our LC extends to $z=0.4$ and covers an on-sky area of $8840 \ \mathrm{deg^2}$ ($\mathrm{RA}\in[120^\circ,250^\circ], \ \mathrm{Dec}\in[0^\circ,68^\circ]$), similar to the SDSS DR7 Legacy footprint of $8423 \ \mathrm{deg^2}$ \citep{Abazajian2009}. This LC contains $\sim17.5$ million galaxies. 

The second step is to generate photometry for the galaxies in our LC. For this we produce a spectral energy distribution (SED) of each galaxy in the LC using ProSpect \citep{Robotham2020} and Viperfish\footnote{Available on GitHub: \href{https://github.com/asgr/Viperfish}{\url{https://github.com/asgr/Viperfish}}. Static version: \citet{viperfish_053}}. ProSpect is a package that combines the stellar population synthesis libraries of \citet{Bruzual2003} or \citet{Vazdekis2016} (chosen by user) with the dust attenuation model of \citet{Charlot2000} and the dust re-emission of \citet{Dale2014}. Viperfish is a wrapper built on top of ProSpect, enabling streamlined extraction of star formation and metallicity histories from the {\tt star\_formation\_histories} file generated by {\sc Shark}. For this work, we generated photometry in the SDSS $g$ and $r$ bands using the \citet{Bruzual2003} templates in ProSpect, adopting the dust parametrization used in \citet{Lagos2019}. More information on LC generation and ProSpect photometry for {\sc Shark} can be found in \citet{Chauhan2019} and \citet{Lagos2019}, respectively.

\begin{figure}
    \includegraphics[width=\linewidth]{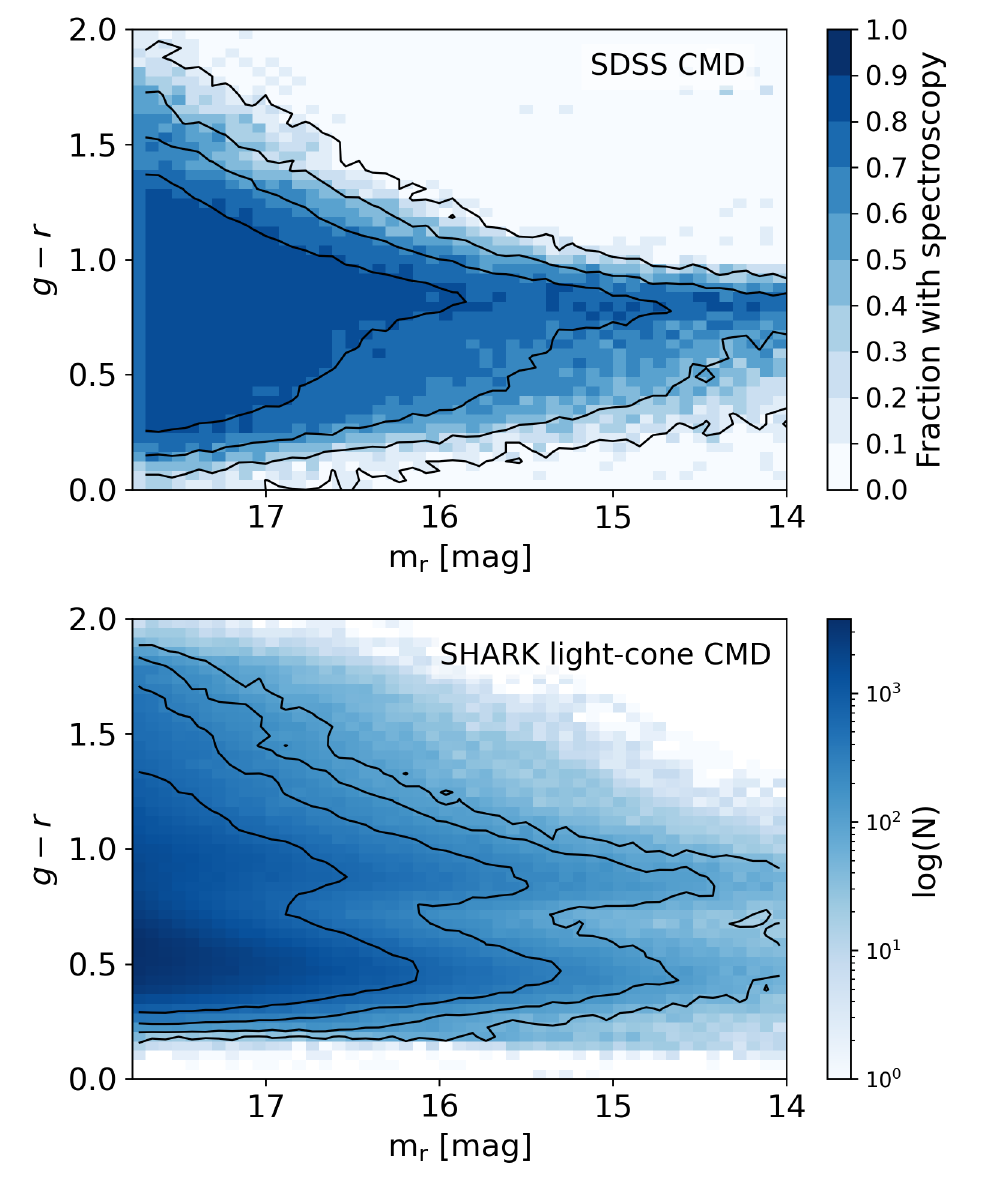}
    \caption{Top: $g-r$ colour magnitude diagram of SDSS galaxies, coloured by the fraction of galaxies within each cell with SDSS spectroscopy. The contours show the $20, 40, 60,$ and $80$th percentiles of the SDSS spectroscopic sample. Bottom: the $g-r$ colour magnitude diagram of SHARK light-cone galaxies before matching to the SDSS, coloured by the number of galaxies in each cell. The contours show the $20, 40, 60,$ and $80$th percentiles after the spectroscopic selection defined in Section \ref{sdss-spectro} (i.e. the \SHARKspec sample).}
    \label{fig:CMD_frac}
\end{figure}

\subsection{Accounting for Observational Limitations}
\label{sdss-spectro}
The SDSS sample used in \citetalias{Oxland2024} consisted of only the subset of SDSS galaxies with spectroscopy. To make a fair comparison between {\sc Shark} and SDSS, we therefore need to select a subset of our {\sc Shark} photometric sample as similar as possible to the SDSS spectroscopic sample.

We first determine the fraction of SDSS-imaged galaxies that have spectroscopy in DR7. The SDSS photometric sample is created by selecting galaxies from the {\tt PhotoObj} table in the CasJobs DR7 database\footnote{\href{https://skyserver.sdss.org/CasJobs/}{\url{https://skyserver.sdss.org/CasJobs/}}} that are primary targets ({\tt mode=1}) flagged as galaxies ({\tt type=3}), with extinction-corrected $r$ band Petrosian magnitudes between 14 and 17.77. We cross match this catalogue to all objects in the {\tt SpecObj} table flagged as galaxies ({\tt specClass=2}) to determine which subset of galaxies also have spectroscopy. We remove luminous red galaxies (LRGs) as they were selected to be fainter and further than the main SDSS flux-limited sample \citep{Eisenstein2001}, as well as quasi-stellar objects (QSOs; \citealt{Shen2011}). We restrict our SDSS sample to be within $\mathrm{RA}\in[120^\circ,250^\circ]$ and $\mathrm{Dec}\in[0^\circ,68^\circ]$, matching the footprint of our {\sc Shark} LC. These cuts result in a total sample of 791812 galaxies with SDSS photometry, of which 559370 have spectroscopy.

\begin{figure}
    \includegraphics[width=\linewidth]{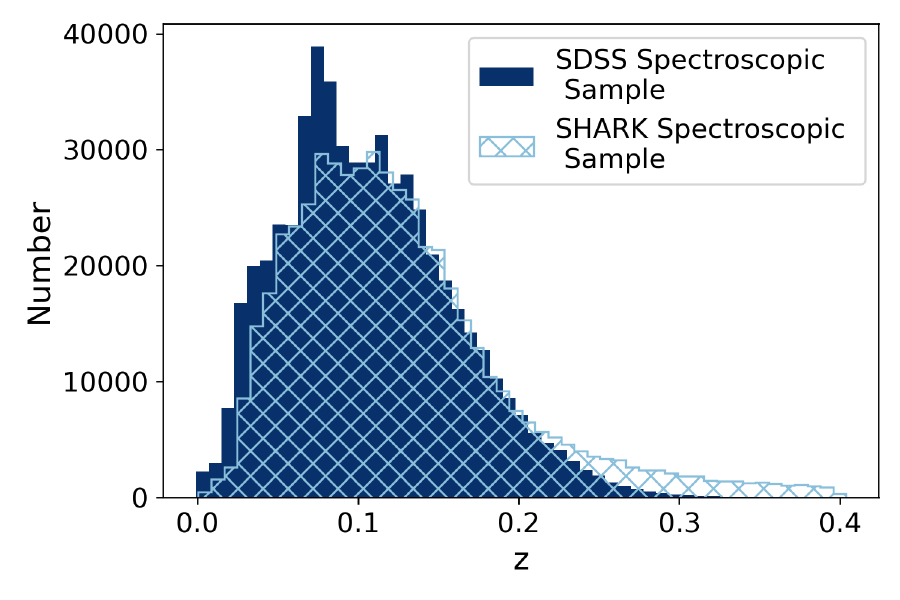}
    \caption{Redshift distribution of SDSS galaxies with spectroscopy (dark blue) and {\sc Shark} galaxies (in light blue) after our CMD selection explained in Section \ref{sdss-spectro}.}
    \label{fig:z_dist}
\end{figure}

To reproduce the SDSS selection in {\sc Shark} we use the fraction of SDSS galaxies with spectroscopy in the $g-r$ colour magnitude diagram (CMD) to define a spectroscopic selection function. The top panel in Figure \ref{fig:CMD_frac} shows the SDSS $g-r$ CMD, with each cell coloured by the fraction of galaxies that have spectroscopy. The contours show the $20, 40, 60,$ and $80$th percentiles of the SDSS spectroscopic sample. We apply this to our {\sc Shark} sample by randomly selecting the same fraction of {\sc Shark} galaxies within each cell of the {\sc Shark} LC CMD (bottom panel in Figure \ref{fig:CMD_frac}) to keep for our analysis. This approach ensures spectroscopic selection is only based on colour and magnitude, with no prior knowledge of redshift required. The resulting {\sc Shark} subsample contains 519582 galaxies, hereafter \SHARKspec, corresponding to the subsample of galaxies in our {\sc Shark} LC that could have spectroscopy (the CMD of this sample is shown by the contours in the bottom panel of Figure \ref{fig:CMD_frac})\footnote{We note that there is no direct colour calibration for {\sc Shark}, which means the galaxy colour distribution in {\sc Shark} is a true prediction and would not be expected to perfectly match the SDSS.}. The redshift distribution of \SHARKspec (Figure \ref{fig:z_dist}) closely resembles that of the SDSS spectroscopic sample, with slightly fewer galaxies at low redshift and slightly more at higher redshift. We only keep galaxies with $z\leq0.1$ when matching to the observations of \citetalias{Oxland2024}.

\subsubsection{Group/Cluster Membership}
\label{membership}
Since our focus in this work is on environmental quenching, we require reliable group membership information. We determine group membership using the {\sc Shark} simulation outputs (including halo mass and virial radius), but identify the group centre and most massive galaxies after applying our spectroscopic selection criteria. This approach more closely mimics the \citet{Yang2007} group finder, avoids relying on the simulation's central/satellite classification, and uses only the subset of galaxies in \SHARKspec when defining group properties. As with the \citet{Yang2007} group catalogue, we retain only groups and clusters with at least three members. The group centre is defined as the luminosity-weighted position of its members, and the most massive galaxy in each halo is flagged as the central. This ensures that our environmental classifications are consistent with the observational methodology, enabling a direct comparison of quenched fractions and quenching timescales between {\sc Shark} and SDSS. While this could lead to some central/satellite contamination, our sample of group/cluster satellite galaxies from \SHARKspec only contains $1.6\%$ {\sc Shark} centrals, so the contamination is low. 

%\begin{equation}
%$    \mu_{50,r} = m_r + 2.5\log_{10}(2\pi R_{50,r}^2)
%$    \label{surface_brightness}
%\end{equation}

\subsubsection{Projected Phase Space Infall Time}
\label{pps}
\begin{figure}
    \includegraphics[width=\linewidth]{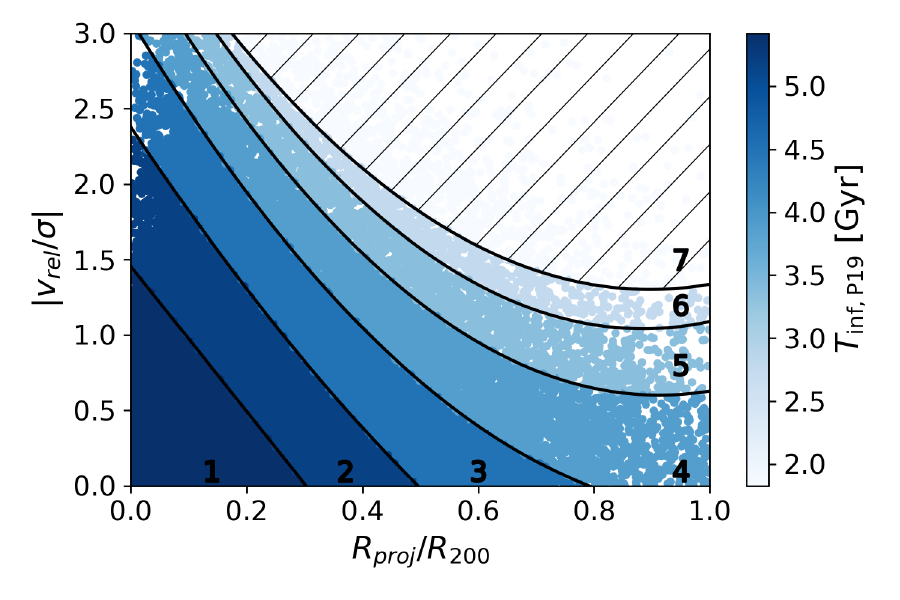}
    \caption{Projected phase space (PPS) distribution of our {\sc Shark} satellite galaxies, used to estimate infall time in Figure \ref{fig:QF_P19}. The 7 different zones correspond to the infall times from \citet{Pasquali2019} and adopted by \citetalias{Oxland2024}. Zone 7, which has a high interloper fraction in observations, is excluded from our analysis.}
    \label{fig:pps}
\end{figure}

\citetalias{Oxland2024} used position in projected phase space (PPS) as a proxy for time since infall (the time since a galaxy first entered its current environment) to study how quickly SDSS satellite galaxies quench. They adopted the ``new zones'' from \citet{Pasquali2019}, who used the hydrodynamical Yonsei Zoom-in Cluster Simulation \citep{Choi2017} to separate PPS into different regions based on satellite galaxy infall time. To compare to \citetalias{Oxland2024}, we use the same zones and infall times for our {\sc Shark} galaxies. The PPS diagram for our \SHARKspec sample is shown in Figure \ref{fig:pps} (due to a high interloper fraction \citep{Reeves2023} and the fact that \citetalias{Oxland2024} added galaxies to region 7 to improve statistics, we do not use any galaxies in zone 7 in our analysis).

Since PPS relies on accurate positional information, we remove orphan galaxies ({\tt type=2}) from our analysis as their positions and velocities are drawn from NFW profiles. This removed 3452 orphans in groups and 3182 orphans in clusters. We confirm that including these orphans do not change our results (see Section \ref{orphans}). Our final sample of {\sc Shark} galaxies in PPS with $\log_{10}(M_{\star}/\mathrm{M}_{\odot}) \geq 9.5$ contains 8512 galaxies in groups and 7636 galaxies in clusters. The field sample consists of 30222 isolated galaxies in the \SHARKspec sample.

We note the number of galaxies quoted in the above section refers to the default {\sc Shark} run where \alphahot and \alphacold are equal to $1$. The procedure is the same for all our other runs with different \alphahot and \alphacold values, and the number of galaxies we obtain for our comparison with SDSS are comparable.

%%%%%%%%%%%%%%%%%%%%%%%%%%%%%%%%%%%%%%%%%%%%%%%%%%

\section{Calibrating hot and cold gas stripping}
\label{sec:calibration}

\begin{figure*}
    \includegraphics[width=\textwidth]{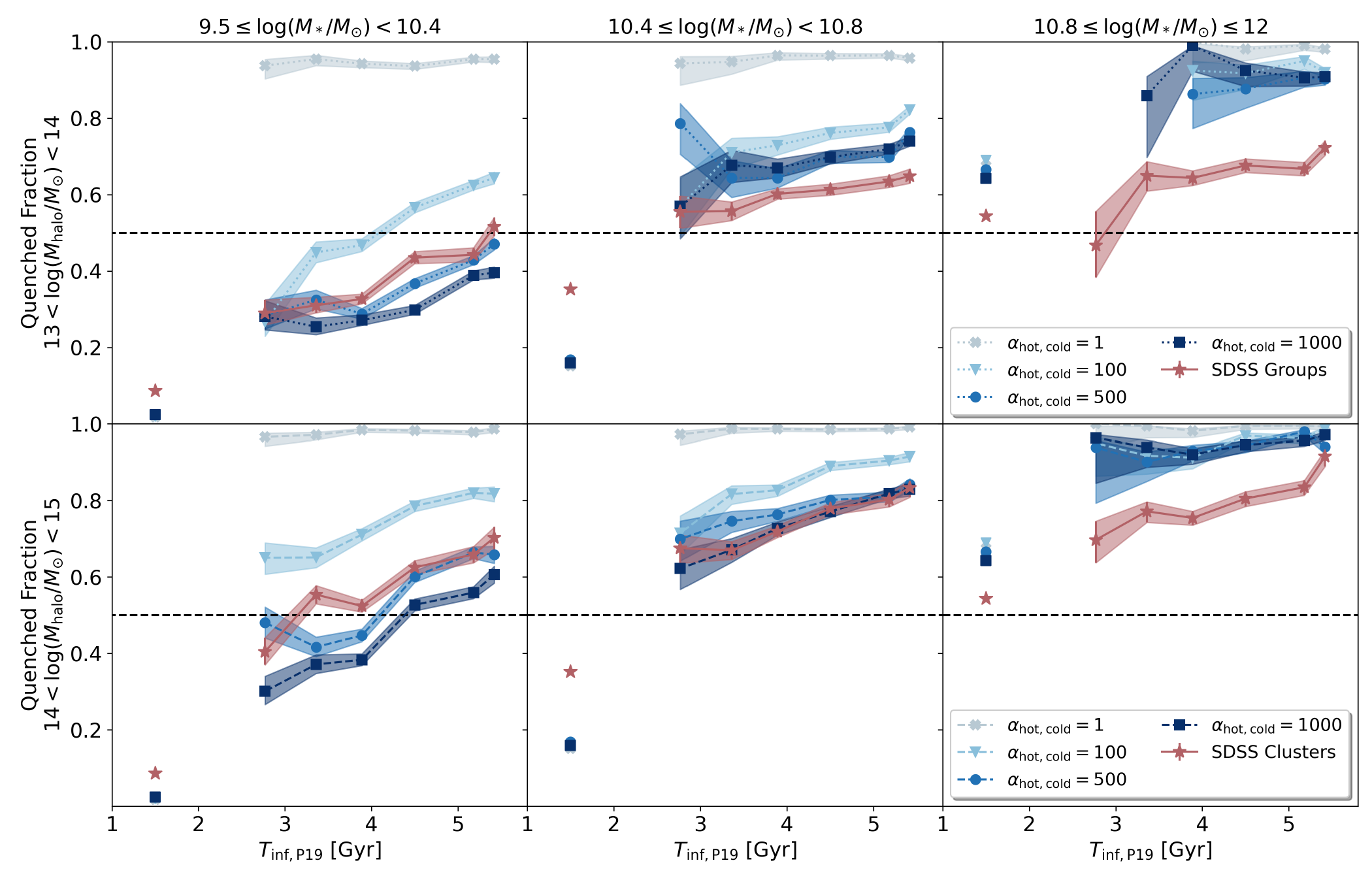}
    \caption{QF of galaxies as a function of time since infall (measured from PPS, see Section \ref{pps}), where the QF is weighted by $1/V_{\mathrm{max}}$. The leftmost panel corresponds to low-mass galaxies, while the middle and rightmost panels contain intermediate- and high-mass galaxies, respectively. The top panel contains galaxies in groups, while the bottom shows galaxies in clusters. Different markers and shades of blue correspond {\sc Shark} runs with different values of \alphahot and \alphacold as shown in the legend. The red stars show the observational results of \citetalias{Oxland2024}. The error bars correspond to the 68 per cent confidence intervals estimated from the beta distribution \citep{Cameron2011}. The single points artificially placed at \TPnineteen = 1.5 Gyr are the field galaxies, and have uncertainties smaller than the marker size. Finally, the black dashed line represents the point at which 50 per cent of galaxies in a given class are quenched. We only show bins with more than 10 galaxies, and exclude galaxies in zone 7 (which have infall times $\lesssim 2.8$ Gyr; see Figure \ref{fig:pps}) due to low numbers and a high interloper fraction in observations}.
    \label{fig:QF_P19}
\end{figure*}

We begin calibrating the stripping prescriptions (Equations \ref{hot_eqn} and \ref{cold_eqn}) by comparing different {\sc Shark} runs with varying values of \alphahot and \alphacold to the observational results of \citetalias{Oxland2024}. Fig. \ref{fig:QF_P19} shows the quenched fraction (QF) as a function of time since infall (\TPnineteen), where \TPnineteen is determined based on galaxy position in PPS (see Section \ref{pps}). We apply a volume correction to our \SHARKspec sample by calculating $V_{\mathrm{max}}$ following the same approach as \citet{Simard2011} (see their Section 3.6), where we weighted our galaxies by 1/$V_{\mathrm{max}}$ (as was done in \citetalias{Oxland2024} to account for the observational incompleteness of the SDSS). The QF is defined in the same way as \citetalias{Oxland2024}:

\begin{equation}
    \mathrm{QF} = \frac{\sum\limits_{i=1}^{N_\mathrm{q}} 1/V_{\mathrm{max},i}}{\sum\limits_{i=1}^{N_\mathrm{total}} 1/V_{\mathrm{max},i}}
    \label{eqn: QF_eq}
\end{equation}

where $N_\mathrm{q}$ is the number of quenched galaxies ($\log_{10}(\mathrm{sSFR}/\rm yr^{-1}) < -11$) and $N_{\mathrm{total}}$ is the total number of galaxies in each PPS bin. The three panels correspond to low ($9.5 \leq\log_{10}(M_{\star}/\mathrm{M}_{\odot}) < 10.4$), intermediate ($10.4 \leq\log_{10}(M_{\star}/\mathrm{M}_{\odot}) < 10.8$), and high ($10.8 \leq\log_{10}(M_{\star}/\mathrm{M}_{\odot}) < 12$) mass galaxies. Galaxies in groups are in the top panel and those in clusters are in the bottom. The QF of field galaxies is plotted at \TPnineteen$=1.5$ Gyr for reference. The results of \citetalias{Oxland2024} are shown by red stars, while the {\sc Shark} results are shown in different markers and shades of blue, corresponding to runs where \alphahot and \alphacold are both equal to 1, 100, 500, or 1000.

The QF generally increases over time for galaxies falling into both group and cluster environments. In the default {\sc Shark} run (\alphahot=\alphacold=1), the QF is systematically too high, exceeding 90 percent across all stellar mass bins. As \alphahot and \alphacold are increased there are fewer quenched galaxies, with the largest effect on the QF occurring for low mass galaxies. This suggests low mass galaxies are the most susceptible to environmental effects, in agreement with previous works \citep[e.g.][]{Wetzel2013, Contini2020, Oxland2024}. The trend of decreasing QF with increasing \alphahot and \alphacold is consistent with the discussion in Section \ref{gas_strip}, where increasing \alphahot and \alphacold increases the hot and cold gas stripping radii, meaning less gas is removed. In panels where the QF always exceeds 0.5, this suggests these galaxies either experienced rapid quenching (within the first ~$3$Gyr which are not plotted in Figure \ref{fig:QF_P19}), and/or they experienced significant pre-processing in a prior group environment.

Qualitatively, the runs with \alphahot and \alphacold both equal to 500 reproduces the SDSS QFs of low and intermediate galaxies very well. The QF of high mass galaxies is higher than the SDSS observations, and changing \alphahot and \alphacold does not have a strong impact on the QF. These galaxies are dominated by AGN feedback quenching, which starts to become efficient when black hole masses are above $\log_{10}(M_{\text{BH}}/\mathrm{M}_{\odot}) \approx 7.5$, corresponding to a stellar mass of $\log_{10}(M_{\star}/\mathrm{M}_{\odot}) \approx 10.5$ \citep{Lagos2024}. Above this mass, environmental quenching gradually becomes less important as higher mass galaxy evolution becomes more tightly linked to AGN feedback.

{\sc Shark} matches the SDSS QF for low and high mass field galaxies relatively well (the difference between the QFs is $\sim0.1$). At intermediate masses however, {\sc Shark} underestimates the number of quenched field galaxies (the difference between the QFs is $\sim0.2$). This is likely due to the fact that this is the well-known transition regime where feedback is just starting to become efficient \citep[e.g.][]{Wetzel2013, Visser2025}. We remind the reader that we follow an SDSS-like central/satellite classification (see Section \ref{sdss-spectro}), so there will be some mixing of true centrals and satellites from matching to the SDSS completeness. Nevertheless, the {\sc Shark} field sample remains dominated by centrals (contains $98.7\%$ centrals, $1.3\%$ satellites), which can still be indirectly affected by environmental stripping. Since centrals grow through the accretion of satellites, the gas content delivered in these mergers depends on how much gas the satellites retain. Satellites that are more strongly stripped contribute less cold gas, and their stripped gas can be locked in the halo by AGN feedback, thereby reducing the fuel for star formation in the accreting central. This effect is visible in the high-mass field sample, where reduced stripping (larger values of \alphahot and \alphacold) corresponds to lower QFs.

While {\sc Shark} uses Particle Swarm Optimization (PSO) for the calibration of many of its free parameters (see Section 3.6 of \citet{Lagos2024}), this approach is not possible for this work. To reduce the computational cost, the PSO calibration in {\sc Shark} is done on a subset of merger trees. In this work, the calibration is performed against a full LC with photometry, which requires the complete galaxy population from the full simulation volume in order to properly populate PPS. In addition, generating photometry is significantly more computationally expensive than running {\sc Shark} (by a factor of $\sim60$). The combined requirements of running the whole simulation volume and photometry for every particle at every optimization step make a PSO approach impractical due to computation costs. Also, other SAMs that have used QFs as calibrators \citep[e.g.][]{Xie2020, Ayromlou2021} typically derive QFs directly from simulation predicted SFRs, and compare these to observed QFs without consideration for selection biases. In contrast, we forward-model the galaxy population to match the survey used for our observational QFs. This approach mitigates observational selection biases and instead places the emphasis on the ability of the model to reproduce the observed data. For this work, we limit our parameter exploration to cases where \alphahot and \alphacold are both set to 1, 100, 500, or 1000. While an infinite number of combinations could be tested, we find that \alphahot = \alphacold = 500 provides the best qualitative match to the \citetalias{Oxland2024} SDSS measurements for low- and intermediate-mass satellite galaxies.

While 500 may seem large, the effect on the gas stripping radii is much smaller and results in larger hot gas stripping radii by a factor of $\sim8$ ($500^{1/3}$) and cold gas stripping radii by a factor of $\sim5$ ($500^{1/4}$; see Section \ref{gas_strip}). We discuss possible reasons why larger stripping radii are favoured in comparison to hydrodynamical simulations in Section \ref{alphas}. Additional runs where \alphahot and \alphacold are varied independently are presented in Appendix \ref{appendix:combos}. 

This analysis demonstrates a marked improvement in the agreement between {\sc Shark} simulations and observations, relative to the results of \citet{Bravo2023}, who used {\sc Shark} v1.0 to compare satellite quenching timescales with those measured in the Galaxy Mass and Assembly Survey (GAMA; \citealt{Driver2011, Driver2022}). Although quenching was defined differently in their study, they applied a consistent definition between simulations and observations, measuring it as the time between a galaxy leaving the blue population and entering the red population. They found that {\sc Shark} systematically under-predicted quenching timescales at low stellar masses, which they attributed to the instantaneous stripping of hot halo gas in {\sc Shark} v1.0 being too aggressive. The transition to gradual stripping of both the hot halo and cold ISM in {\sc Shark} v2.0 appears necessary to reproduce the observed quenching trends for satellite galaxies. 

%%%%%%%%%%%%%%%%%%%%%%%%%%%%%%%%%%%%%%%%%%%%%%%%%%

\section{Quenching Timescales in {\sc Shark}}
\label{sec:satellite_quenching}
\label{quenching timescales}

In Section \ref{sec:calibration} we found that a value of 500 is the best-fit parameter for both \alphahot and \alphacold in order to reproduce the observational results of \citetalias{Oxland2024}. In this section we return to the {\sc Shark} simulation, without applying any observational selections, to understand what {\sc Shark} predicts for the quenching timescales of satellite galaxies, and to assess the relative roles RPS and starvation play in environmental quenching.

\begin{figure*}
    \includegraphics[width=\textwidth]{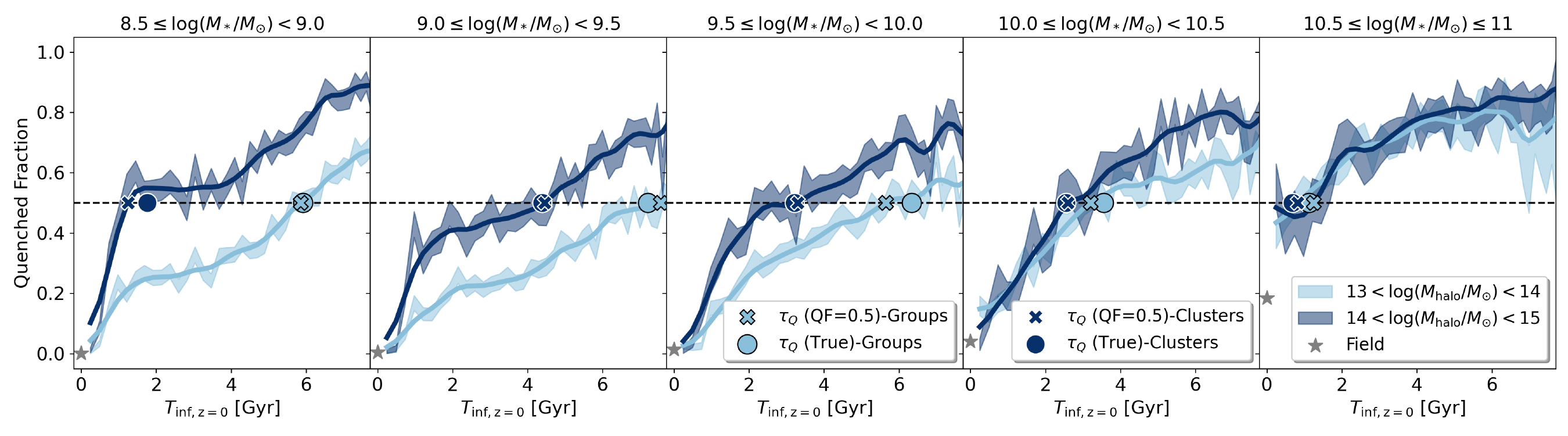}
    \caption{Quenched fraction (QF) as a function of time since infall into the current $z=0$ group/cluster (\TPz), for first infalling satellites with \alphahot = \alphacold = 500. Stellar mass increases from left to right. The light blue and dark blue correspond to galaxies in groups and clusters, respectively. The shaded region shows the 68 per cent confidence intervals on the QF estimated from the beta distribution \citep{Cameron2011}. Grey stars at \TPz = 0 Gyr mark field galaxies, whose uncertainties are smaller than the marker size. Solid lines show the trendlines after applying a Gaussian filter of $1.5\sigma$. Crosses represent the quenching timescales defined when the QF$=0.5$, and circles show the true median quenching timescales (plotted at QF=0.5 for convenience) calculated from the SFHs (see Section \ref{sec:satellite_quenching}). The $95$ percent confidence intervals on the median quenching timescales are smaller than the circular markers in all panels.}
    \label{fig:QF_true}
\end{figure*}

Figure \ref{fig:QF_true} presents the QF as a function of the time since infall into the current ($z=0$) group/cluster for all {\sc SHARK} galaxies (not those in the LC used in Section \ref{sec:calibration}). For this analysis we focus of first infallers, galaxies that became a satellite ({\tt type=1}) when they fell into their current environment and thus are experiencing environmental effects for the first time. Since we are not limited by observational constraints, we now include low mass galaxies down to $\log_{10}(M_{\star}/\rm M_{\odot}) = 8.5$ which are well converged given the mass resolution of the L210N1536 simulation. At the high-mass end, we limit the sample to $\log_{10}(M_{\star}/\rm M_{\odot})\leq 11$ to ensure a sufficient number of star-forming satellites for statistically-robust QF statistics. We note the stellar masses are those at redshift 0. To remove the short-period stochasticity and focus on long-term trends, we apply a $1.5\sigma$ Gaussian smoothing filter to the QF curves (shown by the solid lines plotted over the shaded regions). It is clear that for all stellar masses, the QF increases over time as galaxies fall into group and cluster environments. The QF is systematically higher in clusters compared to groups for satellites with $\log_{10}(M_{\star}/\rm M_{\odot})< 10$, while the trends for galaxies in both environments become more similar at higher masses. As expected, the QF is the lowest in the field where there are no environmental quenching mechanisms at play.

\begin{figure}
    \includegraphics[width=\linewidth]{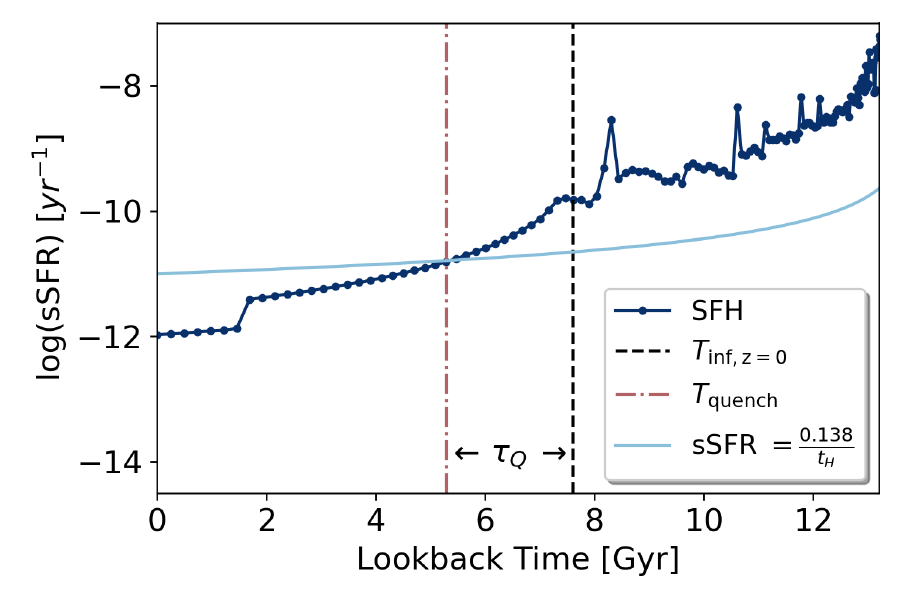}
    \caption{Example star formation history (SFH) for one satellite galaxy in our sample, illustrating how we calculate the quenching timescales described in Section \ref{sec:satellite_quenching}. Dark blue points show the sSFR at each simulation snapshot. The time since infall into the current ($z=0$) group/cluster is shown as the black dashed line. The redshift-dependent sSFR threshold used in calculating the true quenching time is shown by the solid light blue line, and the quenching time (where the SFH crosses this threshold) is shown by the red dot-dashed line. The difference between the infall time and the quenching time is the quenching timescale $\tau_{Q}$.}
    \label{fig:SFH}
\end{figure}

We also show two estimates for quenching timescale ($\tau_Q$) in Figure \ref{fig:QF_true}. The first (plotted as crosses) is determined by the time at which the QF crosses a threshold of 0.5, which defines the time at which the majority of a given population has quenched (as in \citetalias{Oxland2024}). The second is the true median quenching timescale (plotted as circles), calculated from the star formation histories (SFHs) of the satellite galaxies. Here, a galaxy is defined as quenched when its sSFR drops below a redshift-dependent threshold:
\begin{equation}
    \mathrm{sSFR} < 0.138/t_\mathrm{H}(z)
    \label{quenching_threshold}
\end{equation}
where $t_\mathrm{H}(z)=1/H(z)$ and $H(z)$ is Hubble's constant at redshift $z$. Equation \ref{quenching_threshold} corresponds to $\log_{10}(\mathrm{sSFR}/\rm yr^{-1}) < -11$ at $z=0$. This evolving threshold accounts for the redshift evolution of the star-forming main sequence \citep[e.g.][]{Franx2008,Tacchella2019}. The quenching timescale is then the difference between time since infall (\TPz) and the quenching time ($T_\mathrm{quench}$), a common definition used \citep[e.g.][]{Cortese2021, Visser2025}. Figure \ref{fig:SFH} shows an example galaxy SFH and how the associated quenching timescale is calculated.

Many satellite galaxies remain star-forming at $z=0$ so they do not have well-defined quenching timescales. To properly account for these right-censored data points (galaxies for which quenching has not yet occurred), we apply survival analysis, specifically the Kaplan–Meier estimator. Originally developed to estimate the lifespans of individuals in medical statistics \citep{Kaplan1958}, survival analysis is particularly well-suited to this problem, as it allows us to statistically infer the distribution of quenching timescales even when the full quenching event has not been observed for all galaxies. We use the \textsc{lifelines} Python package \citep{davidson_pilon_2024}, applying the method to each stellar and halo mass bin shown in Figure \ref{fig:QF_true}, and plot the resulting median quenching timescales as blue circles. These timescales are broadly consistent with those measured using the $\mathrm{QF}\geq0.5$ definition, providing confidence that the method used in \citetalias{Oxland2024} (namely when the $\mathrm{QF}\geq0.5$) recovers timescales close to the true quenching timescales. For galaxies most influenced by their environment ($\log_{10}(M_{\star}/\mathrm{M}_{\odot}) \lesssim 10.5$), the quenching timescales are $\gtrsim2$Gyr.

We note the similarity between the true median quenching timescales and those estimated from when the $\mathrm{QF}\geq0.5$ in Figure \ref{fig:QF_true} are quite striking. This result suggests if one has a model to accurately predict the infall time of a galaxy, the true median quenching timescale can be recovered remarkably well by the point at which the $\mathrm{QF}\geq0.5$. In practice, however, obtaining accurate infall times from a model is challenging. Methods using PPS, for example, are approximate and fundamentally limited by projection effects and observational uncertainties. Infall times are often estimated using coarse PPS bins calibrated on simulations (as in Figure \ref{fig:pps}) which do not capture the full range of uncertainties in the underlying observables. For this reason, and the fact the stellar mass bins are so different, a detailed comparison between the timescales inferred from Figure \ref{fig:QF_P19} and \ref{fig:QF_true} is non-trivial even though there is a clear conceptual connection.

%%%%%%%%%%%%%%%%%%%%%%%%%%%%%%%%%%%%%%%%%%%%%%%%%%

\section{Discussion}
\label{sec:discussion}
In the previous section, we have calculated the quenching timescales for first-infall {\sc Shark} satellite galaxies, using the best-fit \alphahot and \alphacold parameters found in Section \ref{sec:calibration}. Direct comparisons to quenching timescales presented in other works is difficult due to the numerous ways these timescales are defined \citep[see discussion in ][]{Cortese2021}. We are, however, still able to compare the general trends found, offering insight into the dominant physical processes driving environmental quenching. 

First, we find in Figure \ref{fig:QF_true} that quenching timescales are strongly stellar-mass dependent, with the shortest timescales occurring at both the low mass and high mass ends, with a peak in quenching time around $\log_{10}(M_\star/\rm M_{\odot}) \sim 9-9.5$. Low mass galaxies ($\log_{10}(M_\star/\rm M_{\odot}) \lesssim 9$) are susceptible to rapid environmental quenching due to their shallow potential wells, while high mass galaxies ($\log_{10}(M_\star/\rm M_{\odot}) \gtrsim 10.5$) have lower SFRs at infall, and they continue to experience internally-driven quenching due to AGN feedback after infall.

The longest timescales occur at intermediate stellar masses ($9<\log_{10}(M_\star/\rm M_{\odot})<9.5$). These galaxies have large gas reservoirs that resist rapid quenching, implying a more gradual depletion of their gas supply. The peak in the quenching timescale at intermediate masses is broadly consistent with other studies, where it is thought to roughly align with the minimum feedback efficiency \citep[e.g.][]{Fillingham2015, Davies2019, Davies2025, Visser2025}.

We also find quenching timescales are longer in groups compared to clusters, consistent with expectations from both simulations and observational studies \citep[e.g.][]{Wetzel2013, Oman2021}. In clusters, the denser ICM and higher orbital velocities facilitate stronger ram pressure and therefore shorter quenching timescales compared to groups at fixed stellar mass. Compared with works using the previous version of SHARK, where it was found that the z=0 passive fraction was consistent with observations \citep{Bravo2020} but the quenching timescales were too short and lacked the mass-dependency seen in observations \citep{Bravo2023}, we find that the addition of gradual environmental effects in SHARK v2.0 lead to a significant improvement in the predicted quenching timescales.

Across all environments and stellar masses, the typical quenching timescales exceed $\sim 2$ Gyr, aligning well with a starvation scenario \citep{McGee2009}. The {\sc Shark} model reproduces this behaviour, naturally supporting the interpretation that starvation is the dominant quenching mechanism for the majority of satellites, particularly for galaxies $\log_{10} (M_\star/\rm M_{\odot}) \gtrsim 9$.

\subsection{Physical Meaning of \texorpdfstring{$\alpha_{\mathrm{hot}}$}{ahot} and \texorpdfstring{$\alpha_{\mathrm{cold}}$}{acold}}
\label{alphas}
From the exploration in the Section \ref{sec:calibration}, it is clear \alphahot and \alphacold regulate the strength of hot halo and cold ISM gas stripping where increasing both values decrease the QF, and values of \alphahot = \alphacold = 500 best match the SDSS results of \citetalias{Oxland2024}. These values correspond to increasing the hot halo stripping radii by a factor of $\sim8$ ($500^{1/3}$) and increasing the cold ISM stripping radii by a factor of $\sim5$ ($500^{1/4}$) compared to the radii calculated from the previously adopted \alphahot = \alphacold = 1 run (see Section \ref{gas_strip}). However, the fact that these values are required to bring down the QF of satellite galaxies to more realistic levels begs the question as to what \alphahot and \alphacold could physically represent. While \alphahot and \alphacold could serve to increase the binding energy of the gas, they could equivalently be parameters to decrease the ram pressure felt by each satellite. 

The original hot gas stripping prescription on which {\sc Shark} is based \citep{McCarthy2008} was derived from simulations of relatively massive galaxies falling into massive groups, so the different strength of RPS required may be, in part, due to the diverse satellite population considered in this work. It has also become clear over the years that smoothed particle hydrodynamics (SPH) simulations tend to overestimate gas stripping, especially at the low stellar mass end where galaxies are near the resolution limit \citep{Bahe2015, Bahe2017}. While this numerical effect can likely explain factors of a few compared to the RPS in \citet{McCarthy2008}, the rest of the difference likely comes from the simplified gas geometry that is assumed in SAMs (such as azimuthal symmetry and/or an isothermal hot gaseous halo). In fact some SAMs, such as early versions of L-Galaxies \citep{Guo2011, Henriques2015, Henriques2020}, suppressed all forms of gas stripping in haloes below $\sim10^{14}\mathrm{M}_{\odot}$, otherwise their QFs were too high compared to observations. \alphahot and \alphacold are therefore scaling factors needed to account for numerical effects from SPH simulations, and the simplified geometry assumed in SAMs. The calibrated \alphahot and \alphacold values therefore provide a physically motivated but empirically tuned basis for exploring the quenching timescales of satellite galaxies.

\subsection{Biases and Caveats}
\label{caveats}
When comparing simulations to observations, it is essential to consider the various sources of bias and uncertainty that may influence our results. One key factor is the use of SDSS data in calibrating the {\sc Shark} stripping prescriptions, as described in Section \ref{sec:calibration}. Although SDSS is a widely used survey, its selection function is complex (affected by apparent magnitude limits and spectroscopic targeting biases) which makes it challenging to replicate exactly using simulations. We carefully replicated the SDSS selection function as closely as possible, and note there is little more that can reasonably be done to bring {\sc Shark} into closer alignment with the observations.

One important caveat to keep in mind however is the difference in group-finding algorithms. The SDSS group catalogue used the halo-based \citet{Yang2007} group finder, which differs in both methodology and assumptions from the one used in {\sc Shark}. Differences in how central and satellite galaxies are assigned, as well as the treatment of interlopers (e.g., galaxies misidentified as group members) can lead to discrepancies in inferred group properties such as halo mass and position relative to the group/cluster centre. These projection effects are particularly important in phase-space analyses and may contribute to mismatches in PPS-defined infall times. To minimize these differences, we excluded galaxies in PPS zone 7 (where the interloper fraction is highest) and adopted luminosity-weighted group centres. It is also worth noting the halo definitions inherently differ between {\sc Shark} and SDSS. HBT-HERONS, the merger-tree/halo finder used in {\sc Shark}, defines haloes as bound over-densities tracked consistently across time using all the particles in the simulation. The \citet{Yang2007} group finder defines haloes using position and velocity information of spectroscopically confirmed galaxies, and therefore is subject to all the limitations inherent to observational data. In addition, HBT-HERONS tracks the most bound particles of dark matter subhaloes when finding descendants, and thus is extremely effective at separating structures that could overlap in position (see \citealt{ForouharMoreno2025} for examples). Such structures would be nearly impossible to identify as distinct ones in observations given the way groups are found in projected and redshift space. 

An alternative solution to the issues above is to adopt the strategy of \citet{Bravo2020}, in which the same group finder algorithm was run on mock catalogues. The output not only inherits the limitations of the algorithm applied in observations, but can also serve as a way of determining how clean the resulting group catalogue is. This is beyond the scope of this work, but we highlight that \citet{Bravo2020} found that assuming a fixed fraction of central/satellite contamination gave an equivalent result to the full forward modelling approach above. While the level of central/satellite contamination required for SHARK to match the SDSS \citet{Yang2007} group catalogue is unknown, we tested whether a simple swap of centrals and satellites could reproduce the expected increase of QFs with stellar mass. Introducing a $10\%$ contamination reduced the QFs by $\sim0.1$ in the \alphahot=\alphacold=1 run, but the QFs at low stellar masses remained too high. Even an extreme $50\%$ contamination failed to bring them into agreement. An increase of \alphahot=\alphacold=500 is still required to reproduce the observed trend of increasing QFs with stellar mass.  

\subsection{Effect of Orphans}
\label{orphans}
We exclude orphan galaxies from our analysis since their positions and velocities are drawn from NFW profiles, and therefore their positions in PPS (Figure \ref{fig:pps}) are not accurate. Although orphans only account for $\sim 4\%$ of $\log_{10}(M_\star/\rm M_{\odot}) \geq 8$ galaxies at $z=0$ in {\sc Shark} ($\sim 25\%$ satellites, $\sim 71\%$ centrals), they are skewed towards lower SFRs compared to satellites because they are instantaneously stripped when they become orphans. It is therefore natural to ask if including them would change our results. We confirm that while their inclusion in Figure \ref{fig:QF_P19} slightly increases the QF, the effect is minimal and \alphahot=\alphacold=500 remains the best fit to the SDSS observations of \citetalias{Oxland2024}. 

We also calculate the infall time into the current environment (\TPz), and test whether including first infalling orphan galaxies in Figure \ref{fig:QF_true} change the quenching timescales calculated. While orphans do have lower SFRs than satellites, they also have longer \TPz  compared to satellites. They therefore only increase the QFs slightly at large infall times, but the effect is negligible and does not alter either the derived timescales or our overall conclusions. 

%%%%%%%%%%%%%%%%%%%%%%%%%%%%%%%%%%%%%%%%%%%%%%%%%%

\section{Summary \& Conclusions}
\label{sec:conclusions}
We investigated satellite galaxy quenching in group and cluster environments using the {\sc Shark} semi-analytic model. We calibrated the prescriptions for hot halo and cold ISM gas stripping against recent SDSS observations from \citetalias{Oxland2024}, generating mock SDSS-like data by constructing light-cones and producing photometry for {\sc Shark} galaxies. To replicate the SDSS selection function as closely as possible, we selected a subsample of {\sc Shark} satellites matched to the fraction of SDSS galaxies with spectroscopy in the $g-r$ colour-magnitude diagram. Using projected phase-space position as a proxy for infall time, we compared the quenched fraction as a function of infall time in {\sc Shark} to that in the SDSS. By adjusting the free parameters \alphahot and \alphacold (Equations \ref{hot_eqn} and \ref{cold_eqn}) to match the SDSS results, we found that setting both parameters to 500 reproduces the observed quenching trends for low and intermediate mass galaxies implying reduced stripping efficiency compared to the previous {\sc Shark} default values. In contrast, high-mass galaxies ($\log_{10}(M_{\star}/\mathrm{M}_{\odot}) \gtrsim 10.5$) become increasingly less affected by changes in stripping strength, as their quenching slowly becomes more dominated by AGN feedback.

Applying these calibrated values, we measured the quenching timescales for first-infalling satellite galaxies in {\sc Shark}. Quenching timescales are shorter in clusters compared to groups, decrease with increasing stellar mass above $\log_{10}(M_{\star}/\mathrm{M}_{\odot}) \sim 9$, and typically exceed $\sim2$Gyr, favouring starvation as the dominant mechanism responsible for quenching satellite galaxies.

This work focused on low-redshift satellites, where we have shown {\sc Shark} is able to reproduce observed trends. In future work we will explore satellite quenching at higher redshifts with {\sc Shark}, to gain a more complete picture of galaxy evolution over cosmic time. 

%%%%%%%%%%%%%%%%%%%%%%%%%%%%%%%%%%%%%%%%%%%%%%%%%%

\begin{acknowledgments}
We thank Chris Power for his role in building the SURFS suite and {\'A}ngel Chandro-G{\'o}mez for his work in building the halo and subhalo catalogues from the $N$-body runs. MO and LCP would like to thank the Natural Sciences and Engineering Research Council of Canada for funding. MB is funded by McMaster University through the William and Caroline Herschel Fellowship. This work was supported by resources provided by the Pawsey Supercomputing Research Centre’s Setonix Supercomputer \citep{setonix}, with funding from the Australian Government and the Government of Western Australia.

Funding for the Sloan Digital Sky Survey V has been provided by the Alfred P. Sloan Foundation, the Heising-Simons Foundation, the National Science Foundation, and the Participating Institutions. SDSS acknowledges support and resources from the Center for High-Performance Computing at the University of Utah. SDSS telescopes are located at Apache Point Observatory, funded by the Astrophysical Research Consortium and operated by New Mexico State University, and at Las Campanas Observatory, operated by the Carnegie Institution for Science. The SDSS web site is \url{www.sdss.org}.

SDSS is managed by the Astrophysical Research Consortium for the Participating Institutions of the SDSS Collaboration, including the Carnegie Institution for Science, Chilean National Time Allocation Committee (CNTAC) ratified researchers, Caltech, the Gotham Participation Group, Harvard University, Heidelberg University, The Flatiron Institute, The Johns Hopkins University, L'Ecole polytechnique f\'{e}d\'{e}rale de Lausanne (EPFL), Leibniz-Institut f\"{u}r Astrophysik Potsdam (AIP), Max-Planck-Institut f\"{u}r Astronomie (MPIA Heidelberg), Max-Planck-Institut f\"{u}r Extraterrestrische Physik (MPE), Nanjing University, National Astronomical Observatories of China (NAOC), New Mexico State University, The Ohio State University, Pennsylvania State University, Smithsonian Astrophysical Observatory, Space Telescope Science Institute (STScI), the Stellar Astrophysics Participation Group, Universidad Nacional Aut\'{o}noma de M\'{e}xico, University of Arizona, University of Colorado Boulder, University of Illinois at Urbana-Champaign, University of Toronto, University of Utah, University of Virginia, Yale University, and Yunnan University.  
\end{acknowledgments}

\software{\textsc{python} (\url{https://www.python.org/}), \textsc{astropy} \citep{Astropy2022}, \textsc{matplotlib} \citep{Hunter2007}, \textsc{numpy} \citep{Harris2020}, \textsc{pandas} \citep{McKinney2010}, \textsc{scipy} \citep{Virtanen2020}, \textsc{jupyter} \citep{Granger2021}, \textsc{h5py} \citep{Collette2022}, \textsc{seaborn} \citep{Waskom2021}, \textsc{lifelines} \citep{davidson_pilon_2024}.}

\begin{contribution}
MO was responsible for conducting the research, writing, and submitting this manuscript. LCP provided funding and supervision for the project, while MB provided technical support and guidance. CL provided advice on this project, access to the Pawsey Supercomputing Research Centre, and wrote Section \ref{improvements}. LCP, MB, and CL all edited the manuscript.
\end{contribution}

\appendix
\section{Testing Different Calibration Parameters}
\label{appendix:combos}
\begin{figure*}
    \includegraphics[width=\textwidth]{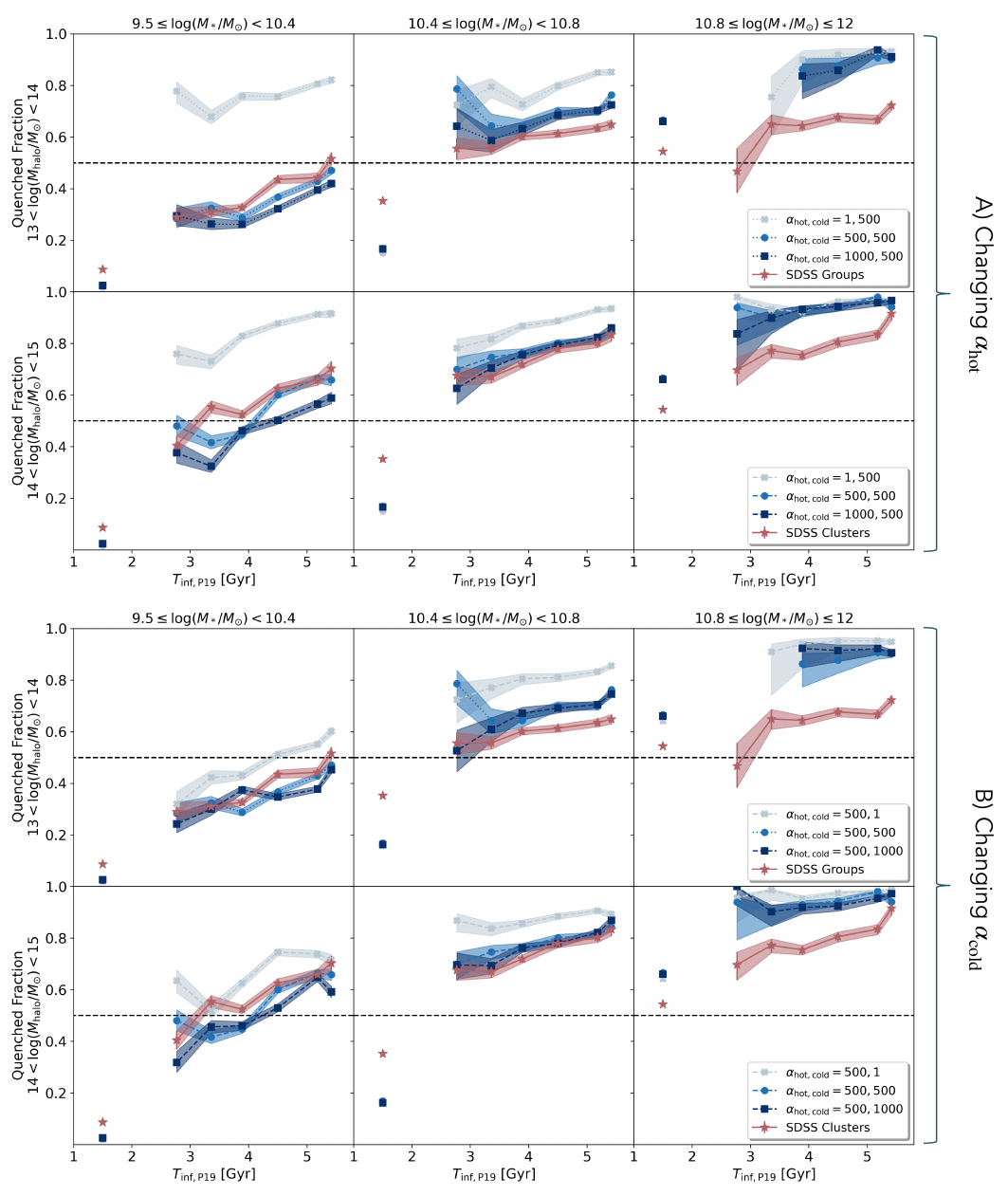}
    \caption{The $1/V_{\mathrm{max}}$ weighted QF of galaxies as a function of time since infall (measured from PPS, see Section \ref{pps}). The separate panels and colours are the same as those defined in Figure \ref{fig:QF_P19}. The top two panels (A) correspond to {\sc Shark} runs where \alphahot varies and \alphacold is kept constant, while the bottom two panels (B) correspond to {\sc Shark} runs where \alphacold varies and \alphahot is kept constant. The different values of \alphahot and \alphacold are in shown by different markers and shades of blue, as shown in the legend. The red stars show the observational results of \citetalias{Oxland2024}. The error bars correspond to the 68 per cent confidence intervals estimated from the beta distribution \citep{Cameron2011}.}
    \label{fig:combos}
\end{figure*}

Due to the computational cost of generating light-cones and photometry with each new {\sc Shark} run, in Section \ref{sec:calibration} we presented the calibration of {\sc Shark} against the SDSS for only four different runs where \alphahot $=$ \alphacold $=1, 100, 500, 1000$. For completeness, in Figure \ref{fig:combos} we present additional {\sc Shark} runs where we independently change \alphahot and \alphacold. The top two panels correspond to the runs where \alphahot varies (1, 500, 1000) and \alphacold is kept at 500, the bottom two panels correspond to runs where \alphacold varies (1, 500, 1000) while \alphahot is kept at 500. 

When \alphahot or \alphacold are varied independently, we observe a systematic decrease in the QF as expected, which is most pronounced among the lowest mass satellites. Modifying \alphahot has a stronger effect on the QF compared to changing \alphacold, as evidenced by the larger variation in the QF in the top two panels. Among all model variations tested, two configurations (1. \alphahot = \alphacold = 500 and 2. \alphahot = 500, \alphacold = 1000) provide the best agreement with the SDSS observations, with only subtle differences between them. Since the case with \alphahot = \alphacold = 500 reproduces the observed trends slightly better at low stellar masses, we adopt it as the preferred parameter set for {\sc Shark}. We note independently changing \alphahot and \alphacold continues to have little effect on high mass galaxies, whose quenching is primarily governed by AGN feedback.  

\bibliography{refs}{}
\bibliographystyle{aasjournalv7}

\end{document}